\documentclass{article}
\usepackage{graphicx}
\usepackage{amssymb,amsfonts,amsmath,stmaryrd,longtable}
\usepackage{cite,enumerate,float,indentfirst}
\usepackage{color}
\hoffset= - 1.2in         % switch off for draft style
\title{Ambiguity in mana and magic definition and knot states}

\author{S. Mironov$^{abcd}$\thanks{e-mail: sa.mironov\_1@physics.msu.ru},
An. Morozov$^{ecd}$\thanks{e-mail: morozov.andrey.a@iitp.ru}}

\date{ }

\begin{document}

\maketitle

\begin{center}
$^a$ {\small {\it INR RAS, Moscow, 117312, Russia}}\\
$^b$ {\small {\it ITMP, MSU, Moscow, 119991, Russia}}\\
$^c$ {\small {\it MIPT, Dolgoprudny, 141701, Russia}}\\
$^d$ {\small {\it Kurchatov Institute, Moscow, 123182, Russia}}\\
$^e$ {\small {\it IITP RAS, Moscow 127994, Russia}}
\end{center}

\begin{abstract}
We study the Mana and Magic for quantum states. They have a standard definition through the Clifford group, which is finite and thus classically computable. We introduce a modified Mana and Magic, which keep their main property of classical computability, while making other states classically computable. We also apply these new definitions to the studies of knot states of 2-strand knots.
\end{abstract}

\section{Introduction}

In recent years papers appeared which discuss magic and mana properties for different models, such as CFT, knot theory and others \cite{ConfMagic,KnotMagic,Chaos}. These quantities characterize how far is the certain quantum mechanical state defined for such a model from the element of the Clifford group \cite{Cliff}. According to the Gottesmann-Knill theorem \cite{GKnill}, Clifford group elements can be effectively modeled on a classical computer. Thus it is claimed that ``magic'' is in effect a non-classicality of a certain state, and mana measures this non-classicality. These properties can be important if one discusses these properties in relation to the quantum computations.

The Gottesman-Knill theorem is based on a fact that Clifford group is a finite subgroup of studied group $G$, which is a tensor product of several $SU(N)$'s. However, it is not the only finite subgroup. One can define infinitely many such subgroups for the same group $G$.  Among these the defining property of the Clifford group is its connection to the sigma matrices. From the point of view of quantum computing there is no need to demand this. Thus depending on the set of problems one wants to present to quantum computer, mana can be defined differently. Our claim is that mana is in fact a relative rather than absolute property.

In the present paper we will present how the Clifford group is usually defined and how it can be modified to get other finite subgroups.
We will apply this new mana definition to studying knot states.

Knot theory is a widely studied subject with lots of relations to other theories. Among others there are connections between knot theory and quantum computations which provide both approaches to calculate knot polynomials using quantum algorithms, as well as describing quantum algorithms as some knot configuration in effective topological field theory \cite{tqc}-\cite{tqcKol2}. This involves calculating knots using unitary matrices through the Reshetikhin-Turaev algorithms \cite{RT}-\cite{RTfin}. Specifically for some particular series of knots any quantum algorithm can be described as consecutive approximations by a series of knots \cite{tqcKol,tqcKol2}.

However, in the present paper we discuss different approach to the knot theory. Mana and magic are properties of the quantum states (density matrices) rather than unitary operations. There is a way to define quantum states, corresponding to a knot \cite{KnotMagic}, using ideas of topological field theories \cite{Ati,Witt}. Matrix elements of this density matrix are made from the knot polynomials in special points. Thus classicality of such states provides us with some information on how can these knot invariants be calculable on classical computer.

Paper is organised as follows. In Chapter 2 we define Clifford group, which is a finite subgroup of the $SU(N)$ group. In Chapter 3 we provide a definition of mana as it is given in other papers on the subject, such as \cite{ConfMagic,KnotMagic,Chaos}. In Chapter 4 we discuss ambiguity in mana definition and show how can the definition be modified to give mana connected to a differen finite subgroup of $SU(N)$. In Chapter 4 we define quantum mechanical states, describing different knots, according to \cite{Ati,Witt,KnotMagic}. In Chapter 5 we study how mana looks like for the knot states and how it can be changed by defining Mana differently.

\section{Clifford group}

Clifford group was first defined by D. Gottesman \cite{Cliff}. Let us take a system of $d$ orthormal states $|k>$, $k=0\ldots d-1$. We take a pair of operators $z$ and $x$:

\begin{equation}
\begin{array}{l}
Z=\sum\limits_{k=0}^{d-1}\omega^k|k><k|,\ \ \ \ \omega=e^{\frac{2\pi i}{d}},
\\
\\
X=\sum\limits_{k=0}^{d-1}|(k+1) \text{mod}\ d><k|.
\end{array}
\label{eq:ZX}
\end{equation}
Using these operators one could define generalized Pauli operators
\begin{equation}
T_{aa^{\prime}}=\left\{\begin{array}{l}
i^{aa^{\prime}}Z^a X^{a^{\prime}},\ \ d=2,
\\
\\
\omega^{-\bar{2}aa^{\prime}}Z^a X^{a^{\prime}},\ \ d>2,
\end{array}\right.
\end{equation}
where $\bar{2}$ is a multiplicative inverse of $2$: $2\times\bar{2}\equiv 1 \text{mod}\ d$. Form the generalized Pauli operators one can defined strings of such operators -- Pauli strings:
\begin{equation}
\label{eq:ps}
T_{\mathbf{a}}=T_{a_1a^{\prime}_1}T_{a_2a^{\prime}_2}\ldots T_{a_na^{\prime}_n}.
\end{equation}

The Clifford group is defined as a set of unitary operators $U$ which transform Pauli string into another Pauli string up to a phase:
\begin{equation}
\label{eq:ClDef}
\mathcal{C}=\left\{U\ \ :\ \ UT_{\vec{a}} U^{\dagger}=e^{i\phi} T_{\vec{b}}\right\}.
\end{equation}

Clifford gates -- elements of the Clifford group -- act in the space of the size $d^n$. There are three Clifford gates which generate the whole group. Two of them act in one $d$-dimensional space, namely phase gate and Hadamard gate:
\begin{equation}
\begin{array}{l}
K=\left(\begin{array}{llllll}
1 \\
& \omega \\
&& \omega^2 \\
&&\ldots \\
&&& \omega^{\frac{(d-1)(d-3)}{2}} \\
&&&& \omega^{\frac{(d-1)(d-2)}{2}}
\end{array}\right),
\\ \\
H=\cfrac{1}{\sqrt{d}}\left(\begin{array}{llll}
1 & 1 & \ldots & 1 \\
1 & \omega & \ldots & \omega^{d-1} \\
\ldots & \ldots & \ldots & \ldots \\
1 & \omega^{d-1} & \ldots & \omega^{(d-1)^2}
\end{array}\right).
\end{array}
\end{equation}

Also there is one operator which acts on a pair of $d$-dimensional spaces:
\begin{equation}
\label{eq:Sop}
S=\sum\limits_{ij}|i;i\oplus j><i; j|.
\end{equation}
In the $SU(2)$ case this operator is the CNOT-gate.

Interesting fact is that there is a finite (although quite large) number of elements in the Clifford group.  Due to this fact anything constructed from the Clifford gates can be effectively simulated on a classical computer in polynomial time. Interesting fact is that Clifford group includes the entangling S-gate (CNOT). Thus non-classicality (magic) is in fact a different parameter from the entanglement.

\section{Mana}

To measure the degree of magic of a certain state a quantity called mana was introduced. It is defined through a phase space point operator $A_{\vec{a}}$:
\begin{equation}
\label{eq:PhP}
A_{\vec{a}}=d^{-n}T_{\vec{a}}\sum\limits_{\vec{b}} T_{\vec{b}}T^{\dagger}_{\vec{a}},
\end{equation}
where $T_{\vec{a}}$ are Pauli strings (\ref{eq:ps}). using these operators one can define discrete Wigner function $W_{\rho}(\vec{a})$ of a state with density matrix $\rho$:
\begin{equation}
W_{\rho}(\vec{a})=\frac{1}{d^n} \text{Tr}\ \rho A_{\vec{a}}.
\end{equation}
Phase space point operators form a complete orthonormal basis in the space of $d^n\times d^n$ matrices, thus density matrix can be constructed from the Wigner functions:
\begin{equation}
\rho=\sum\limits_{\vec{a}}W_{\rho}(\vec{a})A_{\vec{a}}.
\end{equation}
This means that for any physical state when $\text{Tr}\ \rho=1$, set of Wigner functions satisfy $\sum W_{\rho}(\vec{a}) = 1$. Mana is defined as a logarithm of negativity of the set of Wigner functions:
\begin{equation}
M(\rho)=\log \sum\limits_{\vec{a}} |W_{\rho}(\vec{a})|.
\end{equation}

Mana possess several interesting properties which describes why this definition was chosen. First, it can be shown that mana is equal to zero if and only if the density matrix $\rho$ is made from Clifford gates. Second, mana is additive:
\begin{equation}
M(\rho_a\otimes\rho_b)=M(\rho_a)+M(\rho_b).
\end{equation}
Third, mana is related to the second Renyi entropy $S_2$:
\begin{equation}
M(\rho)\leq\frac{1}{2}(L \log d -S_2).
\end{equation}

\section{Ambiguity}

The main property of the Clifford group, related to the ease of classical computation is its finiteness. However Clifford group is not the only finite subgroup of the $U(d)^{\otimes n}$ group. From the Clifford group one can easily construct other finite subgroups using just the rotation matrices.

Let us take as an example the case of one unitary group. Then instead of $X$ and $Z$ operators one can take
\begin{equation}
\tilde{X}=WXW^{\dagger},\ \ \  \tilde{Z}=WZW^{\dagger},
\end{equation}
where $W$ is a matrix from the $SU(d)$ group. %These operators and their products will still of course form a finite group. But they would not satisfy the definition of the Clifford group (\ref{eq:ClDef}).
These provide another set of matrices instead of the generalized Pauli matrices:
\begin{equation}
\tilde{T}=W T W^{\dagger}.
\end{equation}
Instead of Pauli strings one will also get generalized strings which can be rotated independently for each of the generalized Pauli matrices in the string. These generalized Pauli strings will also be related by some finite group:
\begin{equation}
\tilde{\mathcal{C}}=\left\{\tilde{U}\ \ :\ \ \tilde{U}\tilde{T}_{\vec{a}} \tilde{U}^{\dagger}=e^{i\phi} \tilde{T}_{\vec{b}}\right\}.
\end{equation}
These matrices are related to the Clifford group matrices when there is only one $U(d)$ group by rotation with $W$. Thus instead of Hadamard and Phase gate this group will be generated by
\begin{equation}
\tilde{K}=WKW^{\dagger},\ \ \ \tilde{H}=W H W^{\dagger}.
\end{equation}
When tensor product of several $U(d)$ groups is considered, each group can be rotated independently of others. The $S$ operator from (\ref{eq:Sop}) should be modified accordingly:
\begin{equation}
\tilde{S}=\mathcal{W}_2S\mathcal{W}_2^{\dagger},
\end{equation}
where $\mathcal{W}_n=W_1\otimes W_2\otimes\ldots\otimes W_n$, with matrices $W_1$ and $W_2$ acting on the corresponding pair of two dimensional spaces.

By modifying the Clifford group, the definition of Mana also should be modified. Namely phase space point operator instead of (\ref{eq:PhP}) will be equal to
\begin{equation}
\tilde{A}_{\vec{a}}=\mathcal{W}A_{\vec{a}}\mathcal{W}^{\dagger},
\end{equation}
which also modifies the definition of Wigner function:
\begin{equation}
\begin{array}{r}
\tilde{W}_{\rho}(\vec{a})=\frac{1}{d^n}\text{Tr}\ \rho \tilde{A}_{\vec{a}}=\frac{1}{d^n}\text{Tr}\ \rho \mathcal{W}A_{\vec{a}}\mathcal{W}^{\dagger}=
\\ \\
=\frac{1}{d^n}\text{Tr}\ \mathcal{W}^{\dagger}\rho\mathcal{W} A_{\vec{a}}=W_{\mathcal{W}^{\dagger}\rho\mathcal{W}}(\vec{a}).
\end{array}
\end{equation}

\section{Knot states}

Let us discuss applications of the generalized mana definition to the knot theory. To define mana for knots, one should first define some quantum states, related to knots. This can be done as follows, based on the papers by Atiyah \cite{Ati} and Witten \cite{Witt}. From the physics perspective knot theory is a three-dimensional Chern-Simons theory with action
\begin{equation}
S=\frac{k}{4\pi}\int d^3 x \left(\mathcal{A}\wedge d\mathcal{A}+\frac{2}{3}\mathcal{A}\wedge\mathcal{A}\wedge\mathcal{A}\right),
\end{equation}
where $k$ is an integer, called level of the Chern-Simons theory. Knot polynomials are equal to the Wilson-loop averages of the Chern-Simons theory:
\begin{equation}
J^{\mathcal{K}}_r(q)=\left<\text{Tr}_r\oint\limits_{\mathcal{K}} d\vec{x}\vec{\mathcal{A}}\right>.
\end{equation}
Polynomial depends on the gauge group of the Chern-Simons theory, its level $k$, representation $r$ of the gauge group and contour of integration $\mathcal{K}$. In what follows for the sake of simplicity we will speak only of the $SU(2)$ group, when knot polynomials are Jones polynomials. Contour of integration is a closed curve in three-dimensional space or a knot. Finally the answer for the Wilson-loop average happens to be a Laurent polynomial in variable $q$, related to the level of the Chern-Simons theory:
\begin{equation}
q=e^{\cfrac{\pi i}{k+2}}.
\end{equation}
Jones polynomials are in fact related to the representations of the quantum group $U_q(sl(2))$, rather than $SU(2)$ group. When $k$ is an integer then $q$ is a root of integer and then there is a finite number of highest-weight representations of the $U_q(sl(2))$ quantum group. Namely for even $k$ there are only representations, corresponding to the spin $m/2$ for $m\leq k$ \cite{QG}. This allows us to define a basis of representations for the knot states. Namely, state function can be defined as follows:
\begin{equation}
|\mathcal{K}>=\sum\limits_{j=0}^{k-1} J_j |j>.
\end{equation}
Using such a wave function one can also define a density matrix:
\begin{equation}
\rho_{\mathcal{K}}=|\mathcal{K}><\mathcal{K}|,
\label{Pure}
\end{equation}
and study mana for such states. In the present paper we will discuss only two-strand knots, which polynomials can be calculated using the following formula \cite{RT6}:
\begin{equation}
J^{T[2,n]}_j=\sum\limits_{i=0}^j \frac{q^{2j-2i+1}-q^{-2j-2i-1}}{q^{j+1}-q^{-j-1}} \left((-1)^i q^{i^2-2ji-i+j}\right)^{n},
\label{Jones}
\end{equation}
where $n$ is a number of twists in a two-strand braid. For odd $n$ the closure of braid produces a knot, while for even $n$ the result is a link.

\section{Mana for the Knot states}

\begin{figure}[h!]
\centering
\begin{center}
\includegraphics[width=0.45\textwidth]{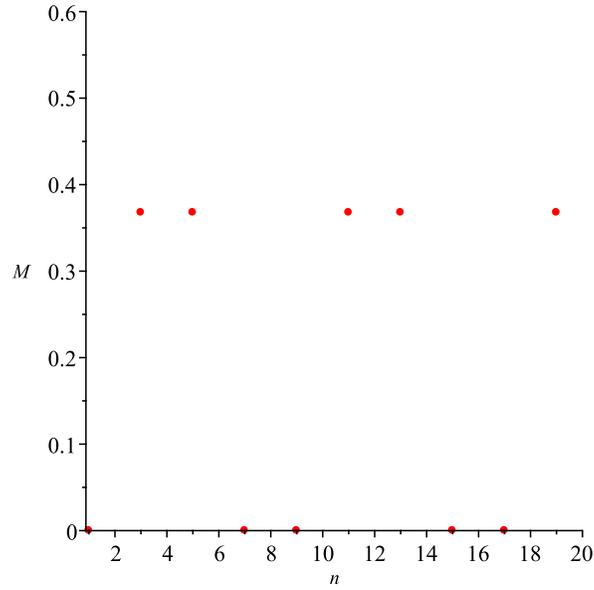}
\end{center}
\caption{1. Mana for two-strand knots for $k=2$ in Clifford group basis \label{f:2Mana}}
\end{figure}

\begin{figure}[h!]
\centering
\begin{center}
\includegraphics[width=0.45\textwidth]{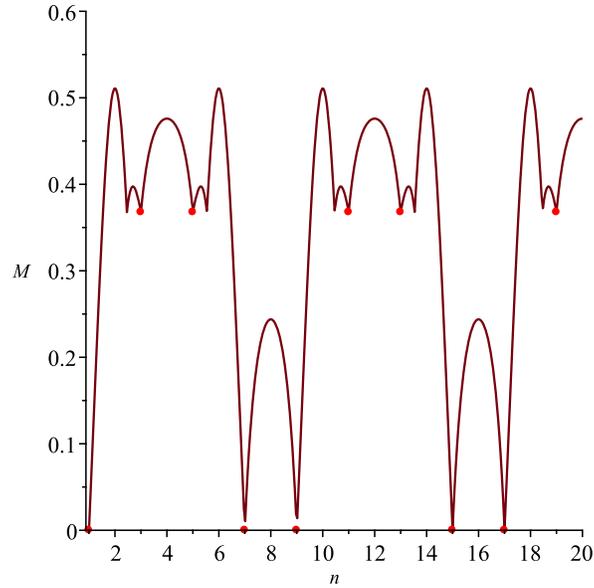}
\end{center}
\caption{2. Mana for two-strand knots for $k=2$ with continuation to arbitrary real $n$ in Clifford group basis.\label{f:2nMana}}
\end{figure}

Mana for two strand knots is periodical in parameter $n$ with period being equal to $k+2$, but besides that mana for many knots coincide. On the Fig.\ref{f:2Mana} mana for $k=2$ is displayed. Knots appear only if $n$ are odd integers in (\ref{Jones}). However, using formula (\ref{Jones}), one can extend this definition to the other values of $n$, which gives graph on the Fig. \ref{f:2nMana}. This is purely an analytical continuation of the results for knots to an arbitrary $n$ by using (\ref{Jones}).

\begin{figure}[h!]
\centering
\begin{center}
\includegraphics[width=0.45\textwidth]{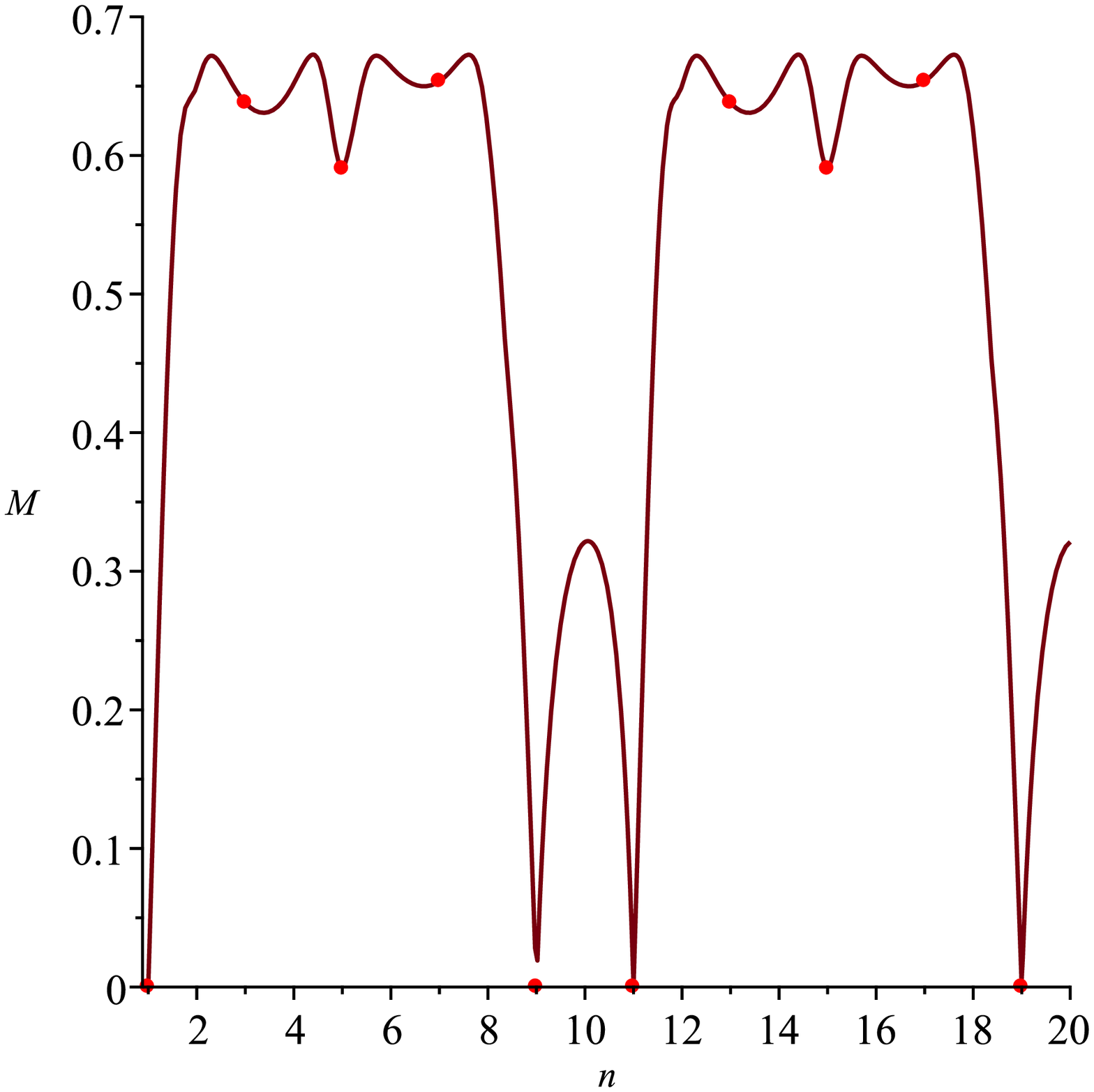}
\end{center}
\caption{3. Mana for two-strand knots for $k=3$  in Clifford group basis. \label{f:3Mana}}
\end{figure}

\begin{figure}[h!]
\centering
\begin{center}
\includegraphics[width=0.45\textwidth]{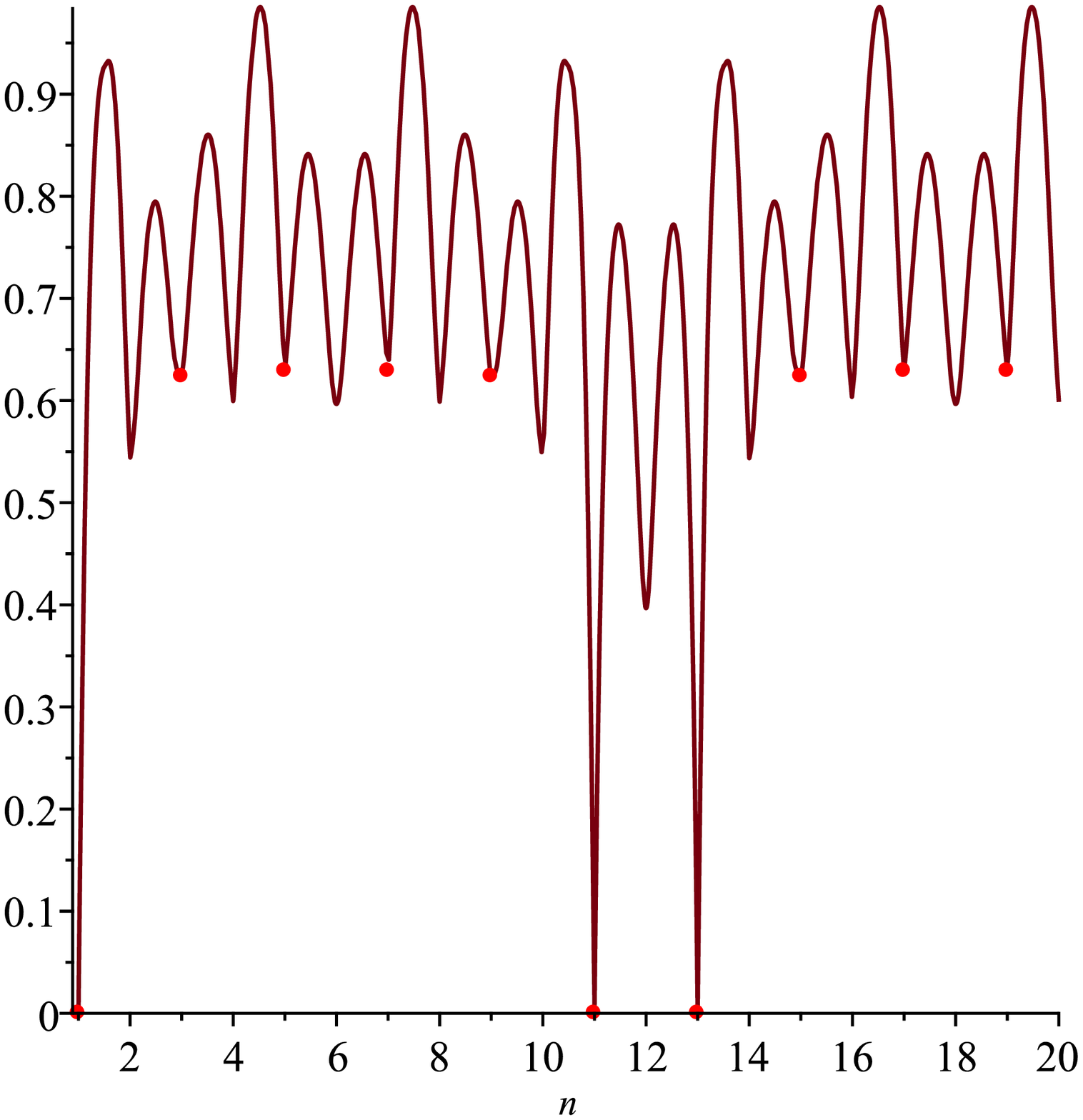}
\end{center}
\caption{4. Mana for two-strand knots for $k=4$ in Clifford group basis.\label{f:4Mana}}
\end{figure}

Similarly mana for $k=3$ (see Fig.\ref{f:3Mana}) and $k=4$ (see Fig.\ref{f:4Mana}) can be produced.

For $k=2$, as can be seen from Fig.\ref{f:2Mana}, there are two distinct values of mana (and in fact only two different density matrices). By changing the basis we calculate mana in, one can make for the other half of 2-strand knots to be equal to zero, while making the mana for the other set to be nonzero. This in fact can be done not only for $k=2$, but also for other cases. This means that by changing the basis one can make different sets of knots states to be effectively calculable on a classical computer.

Specific property of knot states is that they are pure, i.e. $\rho$ is a tensor product of ket and bra vectors (\ref{Pure}). The state for unknot ($n=1$ in (\ref{Jones})) is defined by a vector $v=\frac{1}{\sqrt{d}}[1,1,..,1]$ and $\rho=v^{\dagger}\times v$ for any $k$ and always has mana equals to zero. By transitivity of unitary group it is always possible to rotate any other knot state to the same vector $v=U v_1$. This immediately gives rotated density matrix or, in other words, new basis for Clifford group, $\rho=U v_1^{\dagger} \times v_1U^\dagger=v^{\dagger} \times v$ with zero mana.

In $k=2$ case the only distinguished state besides unknot is trefoil knot ($n=3$ in (\ref{Jones})) with state vector $v_1=\frac{1}{\sqrt{3}}[1,-1,1]$. The easies way to rotate it to the unknot state vector $v$ is by using unitary matrix
\begin{equation}
\setlength{\arraycolsep}{0pt}
S=\left(\begin{array}{ccc}
1\phantom{0} & 0\phantom{0}\phantom{0} & 0
\\
0\phantom{0} & 0\phantom{0}\phantom{0} & 1
\\
0\phantom{0} & -1\phantom{0} & 0
\end{array}\right)
\setlength{\arraycolsep}{5pt}
\label{S2}
\end{equation}.
Of course this matrix is not unique due to stability subgroup SU(2) and U(1) unambiguity in definition of $v$ or $v_1$. But this freedom is not enough to rotate both $v$ and $v_1$ to zero-mana states.
%angle??? between this two vectors is not $\frac{2 \pi}{3}$.%

\begin{figure}[h!]
\centering
\begin{center}
\includegraphics[width=0.45\textwidth]{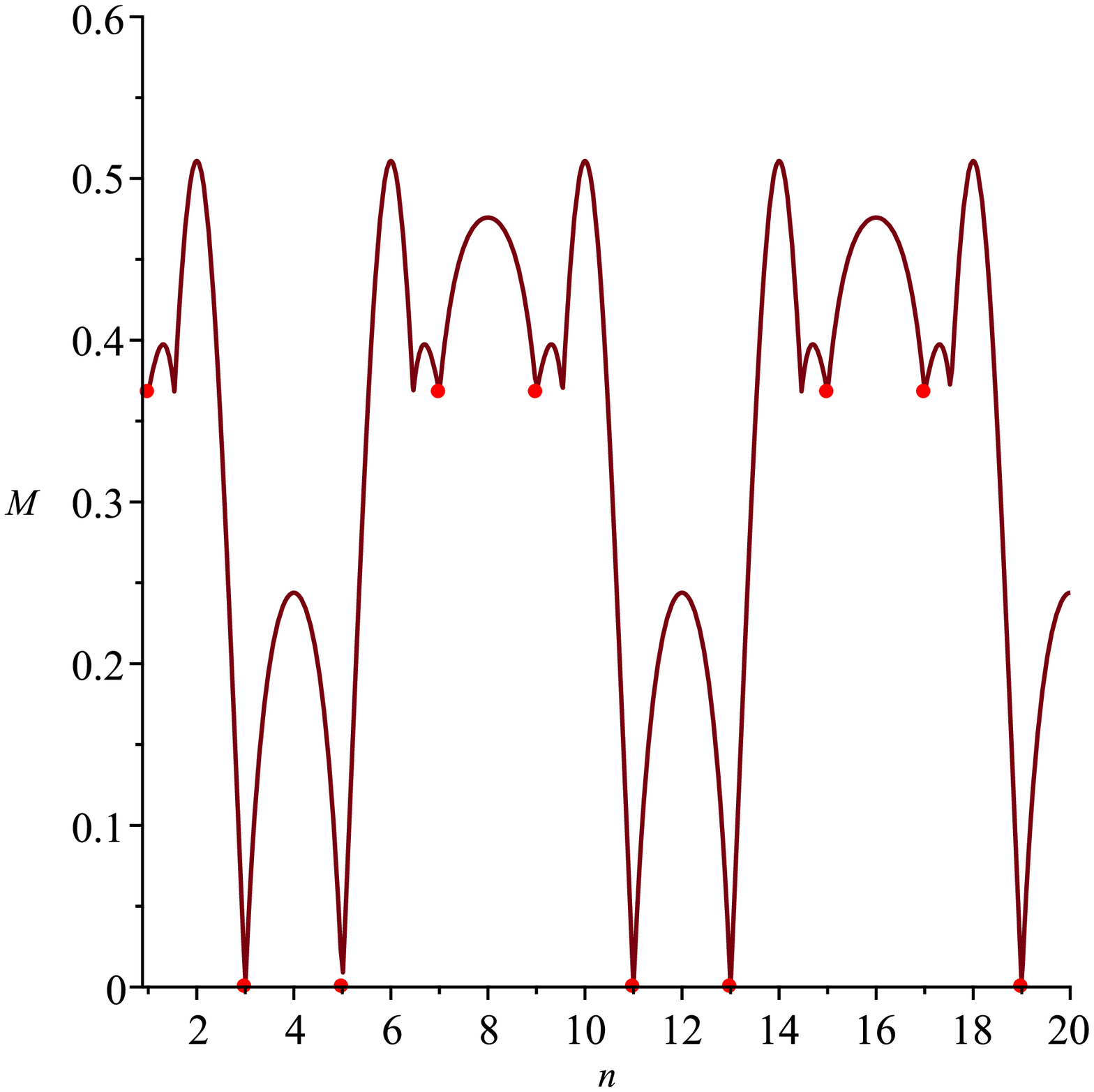}

\end{center}
\caption{5. Mana for two-strand knots for $k=2$ in the basis, rotated by matrix $S$ from (\ref{S2}).\label{f:2ManaR}}
\end{figure}

\begin{figure}[h!]
\centering
\begin{center}
\includegraphics[width=0.45\textwidth]{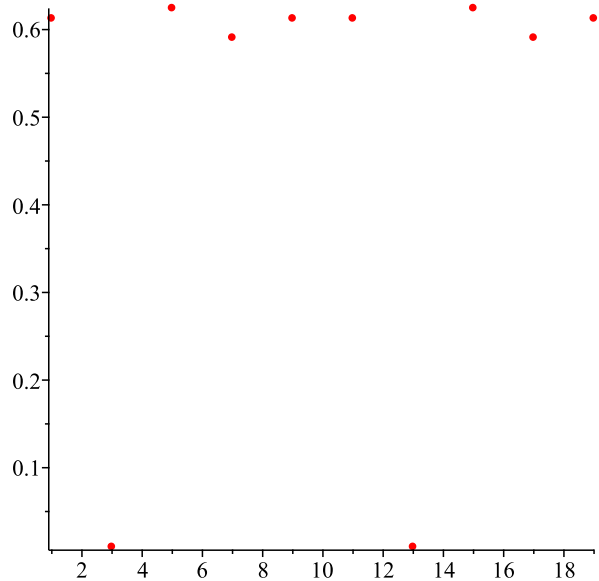}
\end{center}
\caption{6. Mana for two-strand knots for $k=3$ in the basis, rotated by matrix $S$ from (\ref{S3}).\label{f:2ManaR}}
\end{figure}

The $k=3$ case can be analytically solved in a similar way, the rotation matrix between $v$ and $v_1$ can be found for instance in a following way (rounded answer)
\begin{equation}
\setlength{\arraycolsep}{0pt}
S=\left(\begin{array}{cccc}
0.07+0.5 i\phantom{0} & -0.7-0.5 i & 0 & 0
\\
0.7-0.5 i\phantom{0} & 0.07-0.5 i\phantom{0} & 0 & 0
\\
0 & 0 & 0.07-0.5 i\phantom{0} & 0.7-0.5 i
\\
0 & 0 & -0.7-0.5 i\phantom{0} & 0.07+0.5 i
\end{array}\right)
\label{S3}
\setlength{\arraycolsep}{5pt}
\end{equation}

Thus we illustrated for k=2 and k=3 that in different basis different knot states can have zero mana but only separately. There is no basis in which two or more different knot states have zero mana simultaneously. That in fact shows that even if the absolute value of mana for a state is meaningless (it can always be set to zero), the set of values of mana for different states has an invariant property, they can not be set to zero together, thus are not calculable on a classical computer.

%These operators have an important property $x^d=z^d=1$

\section{Conclusion}

In this paper we described how the definition of magic and mana can be changed, while keeping its main property - finiteness of classical calculations. This can be done by rotating either the basis of Clifford group, or rotating the density matrix, which is in principle the same.

We applied these  principles to the studies of mana of knot states. Knots have deep connections with quantum calculations through Reshetikhin-Turaev formalism, which relies on unitary matrices and thus becomes a natural task for quantum computer. Another important connection is topological quantum computer which in turn relies on knots as its basic algorithms.

We studied the mana of 2-strand knots and we also managed to find rotation matrices which allows to change which knots have zero mana and thus are classically calculable.

The goal of this paper was to make it clear that, while mana is an interesting property of the quantum state, which should measure its ``classicality'', its definition is in fact very subjective. It relies on the specific finite subgroup of the $SU(N)$ group. However one can define a different finite subgroup which also changes the mana definition. This new mana also in fact measures classicality, but in a different basis.

Clifford group is only distinguished by its relation to the $X$ and $Z$ operators (\ref{eq:ZX}), which has historical significance, but is not always connected to the real quantum systems, used to build quantum computers. In fact for many quantum computers there is a certain freedom to choose different basic operations, quantum gates. For example for topological quantum computer natural choice of operations are $\mathcal{R}$-matrices, which has no relation whatsoever to the $X$ and $Z$ operators. Thus it is important to understand which states are ``classical'' in relation to the exact choice of basic operations.

We showed using the example of knot states that depending on the basis different knot states can have zero mana in different bases. This in fact means that mana is a relative rather than absolute property. Calculations on a quantum computer (or on a classical reversible computer, which is its classical counterpart) are made by using unitary matrices (or permutation matrices in the classical case), which transform one state of the quantum system into another. If these two states can have zero mana simultaneously in some basis, then the corresponding calculations can be effectively made on a classical computer. Thus mana definition should be chosen specifically for the problem we are trying to study.

While in this paper we used knot states purely as an example to show that mana can be defined differently, it is very interesting to study the dependence of mana in different bases on the knot, as well as mana for the states, related to other knot polynomials. This however still remains to be studied.

\section*{Acknowledgements}

We are grateful for very useful discussions to N. Kolganov.

This work was supported by Russian Science Foundation grant No 18-71-10073.


\begin{thebibliography}{99}

\bibitem{ConfMagic} C.D. White, C. Cao, B. Swingle,
\emph{``Conformal field theories are magical''}, Phys. Rev. B 103, 075145 (2021),
\texttt{arXiv:2007.01303}

\bibitem{KnotMagic} J.R. Fliss,
\emph{``Knots, links, and long-range magic''}, JHEP 04 (2021) 090,
\texttt{arXiv:2011.01962}

\bibitem{Chaos} K. Goto, T. Nasaka, M. Nozaki,
\emph{``Chaos by Magic''},
\texttt{arXiv:2112.14593}

\bibitem{Cliff} D. Gottesman,
\emph{``Theory of fault-tolerant quantum computation''}, Physical Review A. 57 (1): 127–137,
\texttt{arXiv:quant-ph/9702029}

\bibitem{GKnill} D. Gottesman,
\emph{``The Heisenberg Representation of Quantum Computers''}, Group22: Proceedings of the XXII International Colloquium on Group Theoretical Methods in Physics, eds. S. P. Corney, R. Delbourgo, and P. D. Jarvis, pp. 32-43 (Cambridge, MA, International Press, 1999),
\texttt{arXiv:quant-ph/9807006v1}

\bibitem{RT} V.G. Turaev,
\emph{``The Yang-Baxter equation and invariants of links''}, Invent.Math. 92 (1988) 527-553

\bibitem{RT2} N.Yu. Reshetikhin, and V.G. Turaev,
\emph{``Chern-Simons Theory in the Temporal Gauge and Knot Invariants through the Universal Quantum R-Matrix''}, Commun.Math.Phys. 127 (1990) 1-26

\bibitem{RT3} N. Reshetikhin, V.G. Turaev,
\emph{``Invariants of three manifolds via link polynomials and quantum groups''}, Invent.Math. 103 (1991) 547-597

\bibitem{RT4} A. Morozov, and A. Smirnov,
\emph{``Chern-Simons Theory in the Temporal Gauge and Knot Invariants through the Universal Quantum R-Matrix''}, Nucl.Phys. B835 (2010) 284-313,
\texttt{ arXiv:1001.2003}

\bibitem{RT5} A. Smirnov,
\emph{``Notes on Chern-Simons Theory in the Temporal Gauge''},  The Subnuclear Series: Volume 47, The Most Unexpected at LHC and the Status of High Energy Frontier, pp. 489-498 (2011),
\texttt{ arXiv:0910.5011}

\bibitem{RT6} A. Mironov, A. Morozov and An. Morozov,
\emph{``Character expansion for HOMFLY polynomials. II. Fundamental representation. Up to five strands in braid''}, JHEP 03 (2012) 034,
\texttt{ arXiv:1112.2654}

\bibitem{RT7} A. Mironov, A. Morozov and An. Morozov,
\emph{``On Hopf-induced Deformation of Topological Locus''}, Pis’ma v ZhETF, vol. 107, iss. 11, pp. 759 – 760,
\texttt{ arXiv:1804.10231}

\bibitem{RTfin} An. Morozov and A. Sleptsov,
\emph{``New symmetries for the $U_q(sl_N)$ 6-j symbols from the Eigenvalue conjecture ''}, Pis’ma v ZhETF, vol. 108, iss. 10, pp. 721 – 722,
\texttt{ arXiv:1905.01876}

\bibitem{tqc} C. Nayak, S. H. Simon, A. Stern, M. Freedman, S. Das Sarma,
\emph{``Non-Abelian Anyons and Topological Quantum Computation''}, Rev. Mod. Phys. 80, 1083 (2008),
\texttt{arXiv:0707.1889}

\bibitem{tqc1} Yu. Makhlin, S. Backens, A. Shnirman,
\emph{``Two-qubit operation on Majorana qubits in ordinary-qubit chains''}, Pis’ma v ZhETF, vol. 108, iss. 11, pp. 779 – 780,

\bibitem{tqcKauf} L. H. Kauffman, S. J. Lomonaco Jr.,
\emph{``Braiding Operators are Universal Quantum Gates''}, New Journal of Physics,4(2002) 73.1-18;6(2004) 134.1-40,
\texttt{quant-ph/0401090}

\bibitem{tqcus} D. Melnikov, A. Mironov, S. Mironov, A. Morozov, An. Morozov,
\emph{``Towards topological quantum computer''}, Nucl.Phys. B926 (2018) 491-508,
\texttt{ arXiv:1703.00431}

\bibitem{tqcKol} N.Kolganov and An. Morozov,
\emph{``On Hopf-induced Deformation of Topological Locus''}, Pis’ma v ZhETF, vol. 111, iss. 9-10, pp. 623 – 624,
\texttt{ arXiv:2004.07764}

\bibitem{tqcKol2} N. Kolganov, S. Mironov and An. Morozov,
\emph{``Large k topological quantum computer''},
\texttt{ arXiv:2105.03980}

\bibitem{Ati} M. Atiyah,
\emph{``Topological quantum field theories''}, Publications Mathematiques de l’Institut des Hautes Etudes Scientifiques 68 (1988) no. 1, 175–186

\bibitem{Witt} E. Witten,
\emph{``Quantum field theory and the Jones polynomial''}, Commun. Math. Phys.121(1989) 351-399,

\bibitem{QG} B. Abdesselam, D. Arnaudon, A. Chakrabarti,
 \emph{``Representations of $U_q(sl(N))$ at Roots of Unity''}, J. Phys. A: Math. Gen. 28 5495 (1995), \texttt{arXiv:q-alg/9504006}

\end{thebibliography}
\end{document}